\documentclass[pdflatex,sn-mathphys-num]{sn-jnl}

\usepackage{graphicx}%
\usepackage{multirow}%
\usepackage{amsmath,amssymb,amsfonts}%
\usepackage{amsthm}%
\usepackage{mathrsfs}%
\usepackage[title]{appendix}%
\usepackage{xcolor}%
\usepackage{booktabs}%
\usepackage{listings}
\usepackage{float}
\usepackage{caption}
\usepackage{subcaption}

\AtBeginDocument{%
    \setlength{\marginparwidth}{8mm}%
    \setlength{\marginparsep}{3mm}%
    \geometry{reversemp}%
}

\newcommand{\semibold}[1]{\textbf{\textcolor{gray!140}{#1}}}

\theoremstyle{thmstyleone}%
\theoremstyle{thmstyletwo}%

\theoremstyle{thmstylethree}%

\begin{document}

\title[Generative Zero-Shot Environmental Sound Classification]{A Benchmark of Generative Methods for Zero-Shot Environmental Sound Classification}


\author*[1,2]{\fnm{Ysobel} \sur{Sims}}\email{ysobel.sims@mq.edu.au}

\author[1]{\fnm{Alexandre} \sur{Mendes}}

\author[1]{\fnm{Stephan} \sur{Chalup}}

\affil[1]{\orgdiv{School of Computer and Information Sciences}, \orgname{University of Newcastle}, \orgaddress{\city{Callaghan}, \postcode{2308}, \state{NSW}, \country{Australia}}}

\affil[2]{\orgdiv{School of Natural Sciences}, \orgname{Macquarie University}, \orgaddress{\city{Macquarie Park}, \postcode{2109}, \state{NSW}, \country{Australia}}}


\abstract{Zero-shot learning enables models to generalise to unseen classes using semantic information, bridging the gap between training classes and previously unseen test classes. While widely studied in computer vision, its application to environmental audio remains underexplored, and generative approaches have received little attention.

This work presents the first benchmark of generative methods for zero-shot environmental sound classification. Four approaches spanning variational, adversarial, diffusion-based, and denoising paradigms are evaluated. The benchmark includes CADA-VAE and LisGAN, adapted from computer vision, together with two embedding-generation methods introduced in this work: one based on a denoising diffusion probabilistic model (DDPM) and the other on a conditional generative denoising network (CGDN).

Experiments on five environmental audio datasets (ESC-50, ARCA23K-FSD, FSC22, UrbanSound8K, and TAU Urban Acoustic Scenes 2019) and one music dataset (GTZAN) show that generative methods are competitive with established compatibility-based approaches. Among the evaluated generative methods, CGDN achieves the highest average accuracy and is the only one to significantly outperform both the DDPM- and GAN-based methods, while remaining statistically indistinguishable from the strong ALE baseline. These findings suggest that optimisation stability is an important factor in generative zero-shot learning for environmental audio.

Code is provided at \url{https://github.com/ysims/ZeroDiffusion}.}

\keywords{Zero-Shot Learning, Variational Autoencoders, Generative Adversarial Networks, Denoising Models, Diffusion Models, Semantic Embeddings}



\maketitle

\section{Introduction}
\label{sec:intro}

Zero-shot learning (ZSL) is a machine learning paradigm that enables models to recognise instances of previously unseen classes without requiring labelled training examples for those classes~\cite{Larochelle2008ZSL,Lampert2009ZSL}. As machine learning systems are increasingly deployed in domains such as smart cities~\cite{Cugurullo2024}, medicine~\cite{Choy2018318,Ahuja2019,Shah2019}, and robotics~\cite{Valenzuela2024}, they must adapt to new situations and recognise previously unseen concepts~\cite{Enayati2022,Casimiro2022}. ZSL addresses this challenge by enabling models to generalise beyond the classes observed during training.

In ZSL, the classes encountered during testing are not present in the training data. It leverages auxiliary semantic information to transfer knowledge from seen to unseen classes. This capability is valuable in applications where collecting labelled data for every possible class is impractical, including remote sensing~\cite{Avigyan2024}, image classification~\cite{Schonfeld2019,Li2019}, image segmentation~\cite{pastore2021,zheng2021}, environmental audio~\cite{xie2019,Sims2023}, and security~\cite{Barros2022}.

An intuitive way to understand ZSL is through a human analogy. Suppose a person has never seen a cat, bird, or snake but knows semantic descriptions of each animal. When shown a cat, they can identify it from its attributes, such as having four legs and whiskers, while lacking scales or wings. The person uses semantic information from other animals to reason about an unseen class. ZSL methods operate in a similar manner, transferring knowledge from seen classes to unseen classes through auxiliary information.

Practical applications of environmental audio classification include smart cities~\cite{Goulao2024}, detection of traffic noise pollution~\cite{Middya2024}, and hearing aid analysis~\cite{Radha2024}. These applications motivate continued research in environmental audio classification, particularly in the context of ZSL, where comparatively little research has been conducted and performance remains substantially lower than that reported in computer vision.

While generative approaches such as generative adversarial networks (GANs)~\cite{Li2019,Huang2019} and variational autoencoders (VAEs)~\cite{Schonfeld2019,mishra2018} have demonstrated strong performance in zero-shot image classification, their application to zero-shot environmental audio classification remains largely unexplored. Diffusion models have achieved strong results in image generation~\cite{Ho2020,Rambach2022}, but have received little attention in ZSL, particularly outside computer vision.

Six datasets (ESC-50, FSC22, ARCA23K-FSD, TAU Urban Acoustic Scenes 2019, UrbanSound8K, and GTZAN) are used to evaluate five approaches spanning compatibility-based, adversarial, variational, diffusion, and denoising paradigms. These comprise ALE~\cite{xian2019,Akata2015}, LisGAN~\cite{Li2019}, CADA-VAE~\cite{Schonfeld2019}, and two embedding generation methods introduced in this work, based on DDPMs and a Conditional Generative Denoising Network (CGDN). Hereafter, we refer to these zero-shot learning methods simply as DDPM and CGDN.

The contributions of the presented paper are:

\begin{itemize}
    \item A benchmark of generative approaches for zero-shot environmental sound classification, comparing GAN-, VAE-, diffusion-, and denoising-based methods across six audio datasets: ESC-50, FSC22, ARCA23K-FSD, UrbanSound8K, TAU Urban Acoustic Scenes 2019, and GTZAN.
    
    \item The first zero-shot learning benchmark on the FSC22, ARCA23K-FSD, and TAU Urban Acoustic Scenes 2019 datasets.
    
    \item An embedding-space zero-shot learning method based on Denoising Diffusion Probabilistic Models (DDPMs), providing the first evaluation of diffusion-based embedding generation for zero-shot classification.

    \item A Conditional Generative Denoising Network (CGDN)-based embedding generation method for zero-shot environmental sound classification that attains the highest average accuracy across six datasets while remaining the most stable generative method evaluated.
\end{itemize}

\section{Related Work}

This section introduces the terminology used throughout the paper, reviews generative zero-shot learning methods, summarises existing work in zero-shot environmental audio classification, and discusses the datasets considered.

\subsection{Terminology}

This subsection defines the terminology used throughout the paper.

\subsubsection{Embeddings and Auxiliary Data}

Embeddings are vector representations that encode semantic information about complex data, such as images, audio, or text. Formally, an embedding is a vector $x \in \mathbb{R}^{\alpha}$, where $\alpha$ denotes the embedding dimensionality. The methods in this work use deep neural networks to obtain these vector representations from audio data. The embedding network is trained using only data from seen classes to preserve the zero-shot setting.

Embeddings may also be learned from unlabelled text using methods such as GloVe~\cite{pennington2014} and Word2Vec~\cite{mikolav2013}, which encode semantic relationships between words. These embeddings are widely used as auxiliary information in zero-shot learning. A two-dimensional t-SNE~\cite{vandermaaten2008} projection of the Word2Vec embedding space is shown in Figure~\ref{fig:embeddings}. The embeddings for `cat' and `dog' are located closer together than those for `car', illustrating how semantic relationships are preserved in the embedding space.

\begin{figure}[t]
\centering
\includegraphics[width=1.0\linewidth]{EmbeddingsESC50.png}
\caption{A t-SNE graph of Word2Vec word embeddings from ESC-50 dataset class labels. Semantically similar classes are located closer together, such as `cat' and `dog'.}
\label{fig:embeddings}
\end{figure}

ZSL assumes that the identities of unseen classes are known before testing, but no feature examples from those classes are available during training. Each unseen class is instead described using \emph{auxiliary data}. If the identities of the unseen classes are not known in advance, the problem is generally considered open-set recognition rather than zero-shot learning.

The most common form of auxiliary data is a word embedding generated from the class label using methods such as Word2Vec or GloVe. Another common form is manually defined attribute vectors, where each class is represented by properties such as \textit{has fur} or \textit{eats fish}. The Animals with Attributes~\cite{Lampert2009ZSL} and Animals with Attributes 2~\cite{Xian2017} datasets are well-known examples of manually annotated attribute datasets for ZSL research.

\subsubsection{Zero-Shot Learning}

Zero-shot learning involves training a model on a set of seen classes $X$ and evaluating its performance on a disjoint set of unseen classes $Y$, where $X \cap Y = \varnothing$. To preserve the zero-shot setting, examples from unseen classes must not be used during training, including during any pre-training stage.

Many ZSL methods learn a shared semantic space relating feature embeddings to auxiliary class descriptions, enabling transfer from seen to unseen classes. Generative methods instead synthesise feature embeddings for unseen classes from auxiliary information, reducing inference to a conventional supervised classification task.

\subsection{Generative Methods in Zero-Shot Learning}

Generative methods have become an important research direction in zero-shot learning, particularly within computer vision. These approaches use generative models to learn relationships between feature and semantic representations, enabling transfer from seen to unseen classes. Most generative ZSL methods use VAEs and GANs.

\subsubsection{Autoencoders}

Autoencoder-based methods were introduced to zero-shot learning in 2017 by Tsai et al.~\cite{Tsai2017} and Mukherjee et al.~\cite{mukherjee2017}. These approaches used separate autoencoders for visual and semantic modalities and learned a shared latent space between them.

Mishra et al.~\cite{mishra2018} later introduced conditional variational autoencoders, demonstrating successful synthesis of unseen-class feature representations. Schonfeld et al.~\cite{Schonfeld2019} subsequently proposed Cross- and Distribution-Aligned Variational Autoencoders (CADA-VAE), strengthening the shared latent space through cross-alignment and distribution-alignment losses.

\subsubsection{Generative Adversarial Networks}

Generative adversarial networks have also been extensively explored for zero-shot learning. An early example is f-CLSWGAN~\cite{Xian2018ZSL}, which uses a Wasserstein GAN conditioned on semantic information to generate visual features. Wasserstein GANs replace the traditional GAN objective with the Wasserstein distance, improving training stability and reducing mode collapse~\cite{Arjovsky2017}.

Zhu et al.~\cite{Zhu2018} proposed a GAN incorporating a visual pivot regulariser that encourages generated samples to remain close to class-specific feature centroids. Although the approach improved on a basic GAN, the authors observed that unconstrained GAN models performed poorly compared with established ZSL methods.

LisGAN~\cite{Li2019} further improved GAN-based ZSL by conditioning generation on auxiliary data and introducing ``soul samples'', a concept inspired by few-shot learning where representative class examples are used to guide feature generation. The method also includes a self-training stage that incorporates high-confidence unseen samples into the learning process, making it a transductive approach.

Subsequent work addressed limitations of earlier GAN-based methods. GDAN~\cite{Huang2019} introduced a cyclic-consistency loss and semantic regressor, while Task Aligned Generative Meta-learning~\cite{Liu2021_2} combined conditional GANs with meta-learning to better align seen and unseen class distributions. SAGAN~\cite{Tang2022} emphasised improved semantic-visual alignment through classifier training, and ZeroNAS~\cite{Yan2022} applied neural architecture search to automatically discover effective GAN architectures for ZSL.

Despite extensive investigation, GAN-based methods remain challenging to train reliably~\cite{Zhu2018,Yan2022}. Training instability and sensitivity to architecture design continue to limit their practical applicability, particularly in ZSL settings where performance on unseen classes cannot be directly measured during training.

\subsubsection{Diffusion}

Diffusion models have achieved great success in image generation~\cite{Ho2020,Rambach2022}, motivating investigation of their potential for zero-shot learning. However, diffusion-based ZSL remains largely unexplored.

To the authors' knowledge, few works have explored diffusion for zero-shot learning. Li et al.\cite{Li2023} and Clark and Jaini~\cite{clark2023} restrict their methods to computer vision by relying on the pre-trained Stable Diffusion model~\cite{Rambach2022}. While they demonstrate that diffusion-generated information can benefit zero-shot learning, their reliance on large pre-trained models with unknown training data makes it difficult to isolate the contribution of diffusion itself.

Meanwhile, ZeroDiff~\cite{Ye2025} incorporates diffusion mechanisms into a broader generative ZSL framework alongside contrastive and adversarial components. Here, diffusion acts as a data augmentation and representation enhancement module rather than directly synthesising unseen-class features.

Overall, diffusion has primarily been incorporated into composite architectures or large pre-trained systems. Its effectiveness outside computer vision, and as a standalone mechanism for embedding-space generation in zero-shot learning, remains under-characterised.

\subsection{Zero-Shot Environmental Sound Classification}

Zero-shot environmental sound classification remains comparatively underexplored. Islam and Nirjon~\cite{Islam2019} presented one of the earliest studies, proposing a twin neural network architecture evaluated on a ten-class dataset. Xie and Virtanen~\cite{xie2019} later established the first widely adopted benchmark on ESC-50 using an Attribute Label Embedding (ALE)-based compatibility network and introduced the five-fold partition strategies that remain common in subsequent work.

Xie et al.~\cite{xie2021,xie2021_2} subsequently improved performance and highlighted an important consideration for zero-shot evaluation. Large-scale pretrained audio models, such as VGGish trained on AudioSet, may violate the zero-shot assumption when evaluation classes overlap with those encountered during pretraining. In contrast, semantic class embeddings derived from external text corpora do not introduce this issue.

More recently, Wu et al.~\cite{Wu2023} applied the large-scale contrastive audio-text model CLAP to zero-shot environmental sound classification. However, CLAP relies on extensive paired audio-text pretraining, making complete removal of class overlap across common benchmarks difficult and potentially inflating reported zero-shot performance.

Another direction has focused on improving class embeddings rather than the classification model itself. Sims et al.~\cite{Sims2023} enhanced class representations through synonym expansion, demonstrating improved zero-shot performance. The present work adopts this approach for all datasets except ARCA23K-FSD, due to dataset scale, and GTZAN, due to the limited availability of reliable genre synonyms.

\section{Methods}

This section first presents the zero-shot learning formulation used throughout the paper. It then describes the five zero-shot learning methods evaluated in the benchmark.

\subsection{Zero-Shot Learning}

Zero-shot learning is formulated as learning a compatibility function $f(w,z)$ between a feature sample $w$ and a class $z \in X \cup Y$, where $X$ and $Y$ denote disjoint sets of seen and unseen classes, respectively ($X \cap Y = \varnothing$). The model is trained on samples from $X$ and evaluated on samples from $Y$.

The formulation comprises three components. The first is an auxiliary data function $\sigma(z)$ that maps each class label to a semantic representation. In this work, $\sigma(z)$ is implemented using pre-trained Word2Vec embeddings~\cite{mikolav2013}, producing 300-dimensional vectors. Class representations are further refined by averaging embeddings over a set of synonyms following Sims et al.~\cite{Sims2023}.

The second component is a feature embedding function $\tau(w)$ that maps input audio to a semantic feature space. Audio clips are represented as 64-bin log Mel spectrograms, and $\tau(w)$ is implemented using a modified YAMNet trained on seen-class data. The penultimate fully connected layer is used as a 128-dimensional embedding.

The final component is a compatibility function $\xi(\tau(w), \sigma(z))$ that measures similarity between feature and class embeddings. Predictions are obtained by evaluating $\xi$ for all candidate classes and selecting the highest scoring class.

\subsection{CGDN-Based Method}\label{subsec:CGDN}

\begin{figure*}[t]
    \centering
    \includegraphics[width=1.0\linewidth]{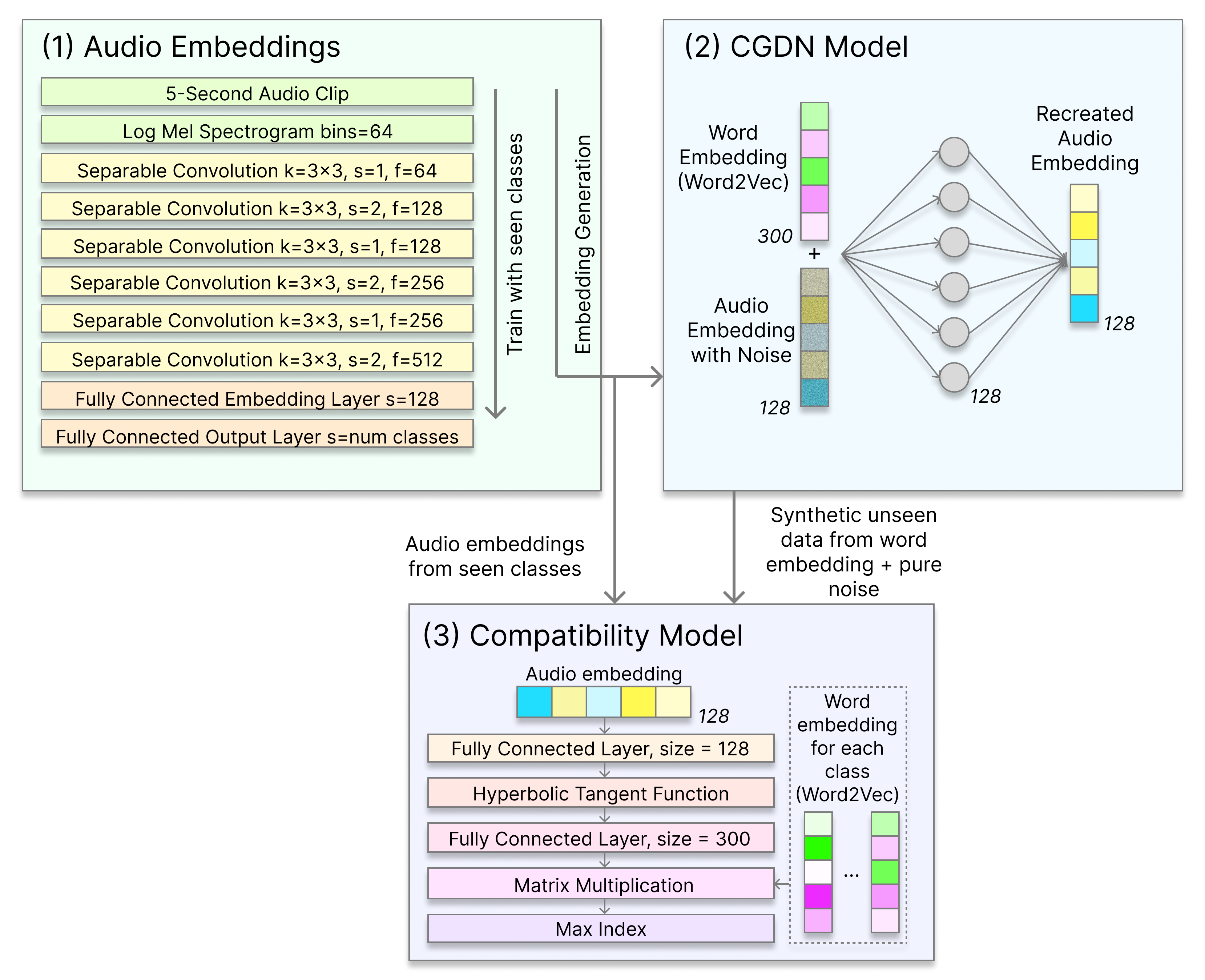}
    \caption{The CGDN-based ZSL method involves training three models. The first (1) is the audio embedding model, used to retrieve audio embeddings. The model is trained with seen classes. Next, (2) CGDN model is trained to produce synthetic embeddings conditioned on class auxiliary data. Finally, (3) a compatibility model is trained using both seen real embeddings and synthetic unseen embeddings. The DDPM-based method works similarly, with the CGDN model and training method replaced with DDPM.}
    \label{fig:diffusion-method}
\end{figure*}

We introduce a zero-shot learning method based on a Conditional Generative Denoising Network (CGDN). The method synthesises feature embeddings for unseen classes by learning to reconstruct clean audio embeddings from corrupted inputs conditioned on the class auxiliary embedding $\sigma(z)$. The synthetic embeddings are combined with seen-class embeddings to train a nonlinear ALE~\cite{xie2021_2} classifier for zero-shot recognition.

CGDN performs single-step conditional denoising by learning to reconstruct clean embeddings from corrupted inputs. During training, a clean embedding $x$ is corrupted by interpolation with a rescaled Gaussian noise vector that preserves its magnitude. Let $\epsilon \sim \mathcal{N}(0, I)$ and $\delta > 0$ be a small constant. The corrupted input is defined as

\begin{equation}
\tilde{x}(p) = (1-p),x + p \cdot \frac{|x|}{|\epsilon| + \delta},\epsilon,
\end{equation}

\noindent where $p \in [0,1]$ increases over training.

The denoiser is conditioned on $\sigma(z)$ by concatenation and trained using a mean-squared error loss between the reconstructed and clean embeddings. Unlike iterative diffusion models, CGDN performs single-step reconstruction, eliminating the stochastic reverse process. This simplifies sampling at the cost of reduced generative flexibility compared with full diffusion models.

Synthetic embeddings are combined with seen-class embeddings to train a nonlinear ALE classifier using a cross-entropy objective. Figure~\ref{fig:diffusion-method} illustrates the CGDN pipeline.

\subsection{DDPM-Based Method}
\label{subsec:DDPM}

We also introduce a zero-shot learning method based on a Denoising Diffusion Probabilistic Model (DDPM). Like the CGDN-based method, it synthesises feature embeddings conditioned on the class auxiliary embedding $\sigma(z)$. However, rather than learning a direct reconstruction mapping, it learns an iterative reverse diffusion process that progressively removes noise from a latent sample.

During training, Gaussian noise is progressively added to a clean feature embedding $x_0$ according to the forward diffusion process. At timestep $t$, the noisy embedding is given by

\begin{equation}
x_t = \sqrt{\bar\alpha_t},x_0 + \sqrt{1-\bar\alpha_t},\epsilon,
\qquad
\epsilon \sim \mathcal{N}(0,I),
\end{equation}

\noindent where $\bar\alpha_t$ is the cumulative product of $(1-\beta_t)$ over timesteps. The denoiser receives the noisy embedding, timestep information, and class auxiliary embedding $\sigma(z)$ as input and is trained to predict the noise component $\epsilon$ using a mean-squared error objective.

The denoising network consists of a compact multilayer perceptron conditioned on both timestep and auxiliary embeddings. Timestep information is encoded using sinusoidal positional embeddings and projected into the network alongside the auxiliary embedding.

During inference, synthetic feature embeddings are generated by iteratively applying the learned reverse diffusion process, beginning from Gaussian noise and conditioning on $\sigma(z)$. The generated embeddings are combined with seen-class embeddings to train a nonlinear ALE classifier.

\subsection{Nonlinear ALE}\label{subsec:ALE}

The compatibility model used throughout this work is the nonlinear Attribute Label Embedding (ALE) network proposed by \citet{xie2021_2}, extending the original Attribute Label Embedding~\cite{Akata2015}. The model serves both as a standalone baseline and as the downstream classifier used by the generative methods evaluated in this work. Synthetic unseen-class embeddings produced by CGDN and DDPM are combined with seen-class embeddings to train the same compatibility model, allowing differences in performance to be attributed to the embedding generation method rather than the classifier.

The model learns the compatibility function

\[
f(w,z)=(B\tanh(Aw))^Tz,
\]

where $w$ is an audio embedding, $z$ is a class embedding, and $A$ and $B$ are trainable weight matrices. Predictions are obtained by evaluating the compatibility score for each candidate class and selecting the highest-scoring class.

\subsection{LisGAN}

LisGAN~\cite{Li2019} is a GAN-based zero-shot learning method that generates feature embeddings conditioned on class embeddings. The method introduces ``soul samples'', representative class features that regularise generation by encouraging synthetic embeddings to remain close to class-specific targets. The original formulation also includes a transductive self-training stage using unlabelled unseen data. Since this work considers the inductive zero-shot setting, this component is omitted.

\subsection{CADA-VAE}

CADA-VAE~\cite{Schonfeld2019} is a variational autoencoder-based zero-shot learning method that learns a shared latent space between semantic and feature embeddings using separate VAEs for each modality. In addition to reconstruction and Kullback-Leibler losses, the method employs cross-alignment and distribution-alignment objectives to encourage both modalities to occupy a common latent space. Synthetic unseen-class embeddings are then generated from semantic embeddings and used to train a classifier.

\section{Experiments}\label{sec:exp}

Experiments are conducted on six audio datasets: ESC-50, FSC22, ARCA23K-FSD, UrbanSound8K, TAU Urban Acoustic Scenes 2019, and GTZAN. The benchmark compares the nonlinear ALE compatibility model~\cite{Akata2015,xie2021_2}, LisGAN~\cite{Li2019}, CADA-VAE~\cite{Schonfeld2019}, the proposed DDPM method, and the proposed CGDN method.

All experiments are repeated ten times, with mean classification accuracy and standard deviation reported. Experiments were conducted on an NVIDIA GeForce RTX 3080 Ti Laptop GPU with 16GB of video memory.

\subsection{Datasets}

The benchmark is evaluated on six audio datasets. Together, they provide a diverse evaluation across curated and large-scale environmental audio, forest acoustics, urban sounds, acoustic scenes, and music genres. Table~\ref{tab:datasets} provides details of the six datasets.

\begin{table}[t]
\centering
\begin{tabular}{lcccc}
\hline
Dataset & Classes & Clips & Duration & Task \\
\hline
ESC-50 & 50 & 2,000 & 5 s & Environmental sound \\
FSC22 & 27 & 2,025 & 5 s & Forest sound \\
ARCA23K-FSD & 70 & 23,170 & 0.3-30 s & Environmental sound \\
UrbanSound8K & 10 & 8,732 & $\leq$4 s & Environmental sound \\
TAU Urban Acoustic Scenes 2019 & 10 & 14,400 & 10 s & Acoustic scene \\
GTZAN & 10 & 1,000 & 30 s & Music genre \\
\hline
\end{tabular}
\caption{Summary of the datasets used in the benchmark.}
\label{tab:datasets}
\end{table}

\textbf{ESC-50}~\cite{piczak2015} is a curated environmental sound classification dataset that is one of the most widely used benchmarks in the domain and forms the basis of previous zero-shot learning evaluations~\cite{xie2019,Sims2023}.

\textbf{ESC-50}~\cite{piczak2015} is a curated environmental sound classification dataset. It is one of the most widely used environmental audio benchmarks and forms the basis of previous zero-shot learning evaluations~\cite{xie2019,Sims2023}.

\textbf{FSC22}~\cite{Bandara2023} is a forest sound classification dataset. Compared with ESC-50, it contains greater background activity and class ambiguity, providing a more challenging evaluation.

\textbf{ARCA23K-FSD}~\cite{Iqbal2021} is a large-scale environmental sound classification dataset derived from FSD50K. Its scale enables evaluation beyond the relatively small datasets commonly used in zero-shot environmental audio classification.

\textbf{UrbanSound8K}~\cite{Salamon2014} contains urban sound classes. The limited number of classes motivates the use of a train-test partition rather than k-fold cross-validation.

\textbf{TAU Urban Acoustic Scenes 2019}~\cite{Mesaros2018} is an acoustic scene classification dataset in which the objective is to identify the recording environment rather than individual sound events. Unlike the Freesound-derived datasets, it was collected directly by the dataset authors.

\textbf{GTZAN}~\cite{Tzanetakis2002} is a music genre classification dataset comprising ten genres. Although outside the environmental audio domain, it provides an additional benchmark for assessing the generality of the evaluated methods.

\begin{table}[htbp]
\centering
\begin{tabular}{l|l|p{5.5cm}}
                                 & \textbf{Partition} & \textbf{Classes}                               \\ \hline
\multirow{5}{*}{\textbf{ESC-50}} & fold 0              & 2, 3, 27, 29, 31, 35, 38, 40, 46, 48          \\ 
                                 & fold 1              & 13, 19, 21, 22, 26, 32, 36, 39, 42, 49        \\ 
                                 & fold 2              & 4, 10, 14, 17, 23, 24, 30, 33, 41, 45         \\ 
                                 & fold 3              & 1, 6, 7, 18, 20, 25, 28, 34, 44, 47           \\ 
                                 & fold 4              & 0, 5, 8, 9, 11, 12, 15, 16, 37, 43            \\ \hline
\multirow{7}{*}{\textbf{ARCA23K-FSD}} & fold 0         & crash cymbal, run, zipper (clothing), acoustic guitar, gong, knock, train, crack, cough, cricket                                                                           \\ 
                                 & fold 1              & electric guitar, chewing mastication, keys jangling, female speech woman speaking, crumpling crinkling, skateboard, computer keyboard, bass guitar, stream, toilet flush                                                           \\ 
                                 & fold 2             & tap, water tap faucet, squeak, snare drum, finger snapping, walk footsteps, meow, rattle (instrument), bowed string instrument, sawing  \\ 
                                 & fold 3             & rattle, slam, whoosh swoosh swish, hammer, fart, harp, coin (dropping), printer, boom, giggle                                      \\ 
                                 & fold 4             & clapping, crushing, livestock farm animals working animals, scissors, writing, wind, crackle, tearing, piano, microwave oven                 \\ 
                                 & fold 5             & trumpet, wind instrument woodwind instrument, child speech kid speaking, drill, thump thud, drawer open close, male speech man speaking, gunshot gunfire, burping eructation, splash splatter                                    \\ 
                                 & fold 6             & female singing, wind chime, dishes pots pans, scratching (performance technique), crying sobbing, waves surf, screaming, bark, camera, organ \\ \hline
\multirow{3}{*}{\textbf{FSC22}}  & train              & 0, 1, 2, 3, 4, 10, 11, 14, 16, 19, 20, 24, 25   \\ 
                                 & val                & 6, 8, 9, 12, 13, 18, 22                         \\ 
                                 & test               & 5, 7, 15, 17, 21, 23, 26                        \\ \hline
\multirow{2}{*}{\textbf{UrbanSound8K}} & train        & 0, 1, 2, 4, 5, 7, 8                             \\ 
                                 & val                & 3, 6, 9                                         \\ \hline
 \multirow{2}{*}{\textbf{TAU 2019}} & train           & 2, 3, 4, 5, 7, 8, 9                             \\ 
                                 & val                & 0, 1, 6                                         \\ \hline
 \multirow{2}{*}{\textbf{GTZAN}} & train              & 0, 1, 2, 6, 7, 8, 9                             \\ 
                                 & val                & 3, 4, 5                                         \\ 

\end{tabular}
    \vspace{10px}
    \caption{Classes in each partition. For ESC-50, random partitions are used from previous work for consistency~\cite{xie2019}. All other datasets have not been benchmarked for zero-shot learning before, so new partitions are created using a random number generator.}
    \label{tab:partitions}
\end{table}

\subsection{Embeddings}

Audio embeddings are extracted using the modified YAMNet architecture proposed by \citet{Sims2023}. The model replaces the original classification head with a 128-dimensional embedding layer followed by a classification layer. For each dataset partition, the network is trained only on samples from the seen classes to preserve the zero-shot formulation.

Across all datasets, the embedding dimension is fixed at 128, the learning rate is $10^{-3}$, and the batch size is 16. Training uses early stopping with a patience of 200 epochs. The resulting embeddings are used as feature representations for all zero-shot learning methods evaluated.

Class embeddings are obtained from the Word2Vec model pre-trained on the Google News corpus\footnote{\url{https://code.google.com/archive/p/word2vec/}}. The same audio and class embeddings are used across all methods to ensure a consistent comparison.

\subsection{Method Configuration}

The benchmark compares the nonlinear ALE compatibility network, CADA-VAE, LisGAN, the proposed DDPM-based method, and the proposed CGDN-based method. Unless otherwise stated, hyperparameters were selected using the validation partitions and then fixed across all datasets. Table~\ref{tab:hyperparameters} summarises the final configurations.

The nonlinear ALE compatibility network~\cite{xie2021_2} is included as a strong compatibility-based baseline. The original hyperparameters were used without modification, except that FSC22 was trained for 30 epochs to improve convergence.

CADA-VAE and LisGAN were implemented from their published descriptions and original implementations. DDPM and CGDN use the architectures described in Sections~\ref{subsec:CGDN} and ~\ref{subsec:DDPM}. For both methods, the number of synthetic embeddings generated for each unseen class equals the average number of samples per seen class. These synthetic embeddings are combined with the real seen-class embeddings to train the nonlinear ALE classifier.

\begin{table}[t]
\centering
\small
\begin{tabular}{p{0.15\textwidth}|p{0.8\textwidth}}
\hline
Method & Configuration \\
\hline
ALE &
Original hyperparameters from~\cite{xie2021_2}; FSC22 trained for 30 epochs. \\
CADA-VAE &
Latent dimension 64; VAE: 80 epochs, learning rate $1.5\times10^{-4}$, batch size 52, MAE reconstruction loss, Adam optimiser. KL weight 1 (warm-up 0--93), cross-reconstruction weight 2.37 (21--75), distribution-alignment weight 5.13 (6--22). Classifier: learning rate $10^{-3}$, batch size 32, Adam, NLL loss; 20--50 epochs depending on dataset. \\
LisGAN &
Generator/discriminator hidden size 128; learning rate $10^{-4}$, batch size 64, Adam, 50 epochs. Classifier: learning rate $10^{-2}$, Adam, NLL loss, 50 epochs. \\
DDPM &
One-hidden-layer MLP (64 units); learning rate $10^{-5}$, weight decay $10^{-4}$, layer normalisation, Adam, 1000 diffusion steps, 200 epochs. ALE classifier: learning rate $10^{-4}$, weight decay $10^{-5}$, batch size 16, Adam, 20 epochs. \\
CGDN &
One-hidden-layer MLP (64 units); learning rate $10^{-3}$, weight decay $10^{-4}$, dropout 0.3, Adam, 80 epochs. ALE classifier identical to DDPM. \\
\hline
\end{tabular}
\caption{Summary of model hyperparameters.}
\label{tab:hyperparameters}
\end{table}

\begin{table*}[t]\small
\centering
\begin{tabular}{ll||lllll}
                                                           &        & \textbf{ALE}        & \textbf{LisGAN}     & \textbf{CADA-VAE}              & \textbf{DDPM}              & \textbf{CGDN}       \\ \hline \hline
\parbox[t]{2mm}{\multirow{5}{*}{\rotatebox[origin=c]{90}{\textbf{ESC-50}}}}      & fold 0 & $0.2278 \pm 0.0091$   & $\semibold{0.2605} \pm 0.0227$   & $\textbf{0.2915} \pm 0.0180$          & $0.2270 \pm 0.0192$          & $0.2505 \pm 0.0073$ \\
\multicolumn{1}{l}{}                                      & fold 1 & $\textbf{0.4932} \pm 0.0093$   & $0.4039 \pm 0.0220$   & $0.3980 \pm 0.0181$          & $0.3605 \pm 0.0493$          & $\semibold{0.4440} \pm 0.0158$ \\
\multicolumn{1}{l}{}                                      & fold 2 & $\textbf{0.2937} \pm 0.0129$   & $0.2582 \pm 0.0187$   & $0.2500 \pm 0.0251$          & $\semibold{0.2908} \pm 0.0513$          & $0.2728 \pm 0.0084$ \\
\multicolumn{1}{l}{}                                      & fold 3 & $0.2680 \pm 0.0117$   & $0.2150 \pm 0.0195$   & $\semibold{0.3013} \pm 0.0251$          & $0.2428 \pm 0.0351$          & $\textbf{0.3363} \pm 0.0108$ \\
\multicolumn{1}{l}{}                                      & test   & $0.2340 \pm 0.0167$    & $\semibold{0.2448} \pm 0.0165$   & $0.2095 \pm 0.0337$          & $0.2238 \pm 0.0394$          & $\textbf{0.2740} \pm 0.0110$ \\ \hline
\parbox[t]{2mm}{\multirow{7}{*}{\rotatebox[origin=c]{90}{\textbf{ARCA23K-FSD}}}} & fold 0 & $\semibold{0.2420} \pm 0.0032$   & $0.2183 \pm 0.0128$   & $0.2398 \pm 0.0125$          & $0.2415 \pm 0.0276$          & $\textbf{0.2830} \pm 0.0074$ \\
\multicolumn{1}{l}{}                                      & fold 1 & $0.2876 \pm 0.0028$   & $0.3059 \pm 0.0107$   & $\semibold{0.3142} \pm 0.0149$          & $0.2912 \pm 0.0377$          & $\textbf{0.3300} \pm 0.0077$ \\
\multicolumn{1}{l}{}                                      & fold 2 & $\semibold{0.3723} \pm 0.0061$   & $0.3086 \pm 0.0121$   & $0.3164 \pm 0.0120$          & $0.3318 \pm 0.0919$          & $\textbf{0.5114} \pm 0.0101$ \\
\multicolumn{1}{l}{}                                      & fold 3 & $\textbf{0.2546} \pm 0.0029$   & $\semibold{0.2113} \pm 0.0148$   & $0.2036 \pm 0.0101$          & $0.1362 \pm 0.0234$          & $0.1618 \pm 0.0103$ \\
\multicolumn{1}{l}{}                                      & fold 4 & $0.2283 \pm 0.0020$   & $0.2173 \pm 0.0113$   & $0.2126 \pm 0.0140$          & $\semibold{0.2625} \pm 0.0513$          & $\textbf{0.3248} \pm 0.0084$ \\
\multicolumn{1}{l}{}                                      & fold 5 & $0.3205 \pm 0.0050$   & $0.3196 \pm 0.0142$   & $0.3567 \pm 0.0144$          & $\textbf{0.4694} \pm 0.0530$          & $\semibold{0.3971} \pm 0.0144$ \\
\multicolumn{1}{l}{}                                      & test   & $\semibold{0.2129} \pm 0.0051$   & $0.1772 \pm 0.0068$   & $\textbf{0.2414} \pm 0.0145$          & $0.1654 \pm 0.0209$          & $\semibold{0.2179} \pm 0.0088$ \\ \hline
\multicolumn{1}{l}{\multirow{2}{*}{\textbf{FSC22}}}       & val    & $0.2650 \pm 0.0088$    & $0.2990 \pm 0.0219$    & $\textbf{0.3179} \pm 0.0137$          & $0.2469 \pm 0.0648$          & $\semibold{0.3044} \pm 0.0069$ \\
\multicolumn{1}{l}{}                                      & test   & $\semibold{0.2968} \pm 0.0070$   & $0.2142 \pm 0.0416$   & $\textbf{0.3290} \pm 0.0339$          & $0.2368 \pm 0.0542$          & $0.2417 \pm 0.0096$ \\ \hline
\multicolumn{2}{l||}{\textbf{UrbanSound8k}}                        & $\textbf{0.4997} \pm 0.0071$   & $0.3848 \pm 0.0238$   & $\semibold{0.4741} \pm 0.0808$          & $0.3722 \pm 0.0787$          & $0.4445 \pm 0.0035$ \\ \hline
\multicolumn{2}{l||}{\textbf{TAU2019}}                             & $0.4377 \pm 0.0040$    & $0.3878 \pm 0.0188$   & $0.3158 \pm 0.0248$          & $\semibold{0.4419} \pm 0.0627$          & $\textbf{0.4857} \pm 0.0395$ \\ \hline
\multicolumn{2}{l||}{\textbf{GTZAN}}                               & $\semibold{0.5705} \pm 0.0109$   & $\textbf{0.5824} \pm 0.0287$   & $0.5650 \pm 0.0463$          & $0.4117 \pm 0.0683$          & $0.5381 \pm 0.0109$ \\ \hline

\multicolumn{2}{l||}{\textbf{Average}} & $\semibold{0.3227} \pm 0.0076$   & $0.2936 \pm 0.0187$   & $0.3137 \pm 0.0242$ & $0.2913 \pm 0.0488$ & $\textbf{0.3422} \pm 0.0112$
\end{tabular}
\vspace{8px}
\caption{Average accuracy and standard deviation over ten random seeds for the non-linear ALE baseline and four generative zero-shot learning methods across six datasets. CGDN achieves the highest overall average accuracy across datasets, although this average is statistically indistinguishable from the ALE baseline (Section~\ref{subsec:stats}). At the dataset level, performance varies across methods, with different approaches achieving the best results depending on the dataset and split configuration. The best result in each row is highlighted in bold, and the second-best is shown in grey. The final row reports the macro-average across datasets. Variance differs across methods, with DDPM exhibiting the highest instability across runs, while ALE shows comparatively lower variance.}
\label{tab:results}
\end{table*}

\section{Results}

Table~\ref{tab:results} presents the results for the five methods: nonlinear ALE, LisGAN, CADA-VAE, DDPM, and CGDN across ESC-50, ARCA23K-FSD, FSC22, UrbanSound8k, TAU 2019 and GTZAN.

Across datasets, performance differences between methods are generally small, with rankings varying by dataset and split configuration. CGDN achieves the highest overall mean accuracy across all datasets. At the dataset level, it is consistently among the strongest methods, attaining multiple top and second-best results across ESC-50 and ARCA23K-FSD, and attains top performance on TAU 2019. On FSC22, CADA-VAE achieves the strongest performance across both validation and test splits, outperforming the other methods consistently on this dataset.

A notable pattern across results is the relationship between variance and performance stability. ALE remains a consistently stable baseline with comparatively low variance, while DDPM exhibits substantially higher variance across runs. CGDN, LisGAN and CADA-VAE sit between these extremes, with CGDN showing more stable behaviour while maintaining higher average performance.

These results indicate that simpler embedding-space generative approaches can remain competitive with more complex generative formulations. In particular, CGDN, which uses a direct conditional denoising objective in embedding space, achieves strong average performance without the instability observed in multi-step diffusion sampling. At the same time, compatibility-based methods such as ALE remain competitive across several datasets, suggesting that gains from generative modelling are strongly coupled to training stability and variance rather than model complexity.

\subsection{Robustness}

Performance varies considerably across both dataset partitions and random seeds. On ESC-50 and ARCA23K-FSD, the relative ranking of methods changes across evaluation folds, indicating that zero-shot performance is sensitive to the choice of seen and unseen class partition. For example, CGDN achieves an average accuracy of 0.5114 on fold 2 of ARCA23K-FSD but only 0.1618 on fold 3, demonstrating that some unseen class partitions are substantially more difficult than others. Similarly, the strongest-performing method differs between partitions, suggesting that performance on a single split is not necessarily representative of overall zero-shot performance.

Performance also varies across repeated training runs. Standard deviation over ten random seeds shows that DDPM exhibits the highest variability across datasets, while ALE remains the most stable method. CGDN demonstrates substantially lower variance than DDPM while achieving the highest average accuracy across all datasets. CADA-VAE and LisGAN generally exhibit intermediate levels of variability.

Together, these results suggest that robustness is an important consideration for zero-shot learning. Although the identities of the unseen classes are known during model development, no labelled examples are available for evaluation or model selection. Methods should therefore perform consistently across different seen/unseen class partitions and repeated training runs. The results indicate that the single-step denoising formulation of CGDN provides a favourable balance between performance and stability, whereas the iterative diffusion process used by DDPM exhibits substantially greater variability.

\subsection{Statistical Analysis}\label{subsec:stats}

To assess whether performance differences between methods are statistically meaningful, a Friedman omnibus test is applied across all methods over per-split accuracy values, following \citet{Demsar2006}. The Friedman test indicates a significant difference between methods ($\chi^2 = 13.51$, $p = 0.009$). Mean ranks are shown in Table~\ref{tab:stats}.

\begin{table}[t]
\centering
\small
\begin{tabular}{lccc}
\hline
Method & Mean rank & $p_{\text{Holm}}$ \\
\hline
CGDN     & 2.06 & --    \\
ALE      & 2.59 & 0.414 \\
CADA-VAE & 3.00 & 0.414 \\
LisGAN   & 3.59 & 0.020 \\
DDPM     & 3.76 & 0.007 \\
\hline
\end{tabular}
\vspace{1.0em}
\captionsetup{width=0.95\textwidth}
\caption{Statistical comparison of the evaluated methods. Mean ranks are obtained from the Friedman test (lower is better). Pairwise comparisons are Wilcoxon signed-rank tests with Holm correction using CGDN as the reference method.}
\label{tab:stats}
\end{table}

Post-hoc Wilcoxon signed-rank tests with Holm correction indicate that CGDN significantly outperforms LisGAN and DDPM, while differences between CGDN and ALE and between CGDN and CADA-VAE are not statistically significant (Table~\ref{tab:stats}).

These results indicate that CGDN significantly outperforms the GAN- and diffusion-based baselines, while remaining statistically comparable to the strongest compatibility-based and VAE-based methods.

\subsection{Embedding Visualisation}\label{subsec:tsne}

\begin{figure*}[t]
    \centering

    \begin{subfigure}[t]{0.32\linewidth}
        \centering
        \includegraphics[width=\linewidth]{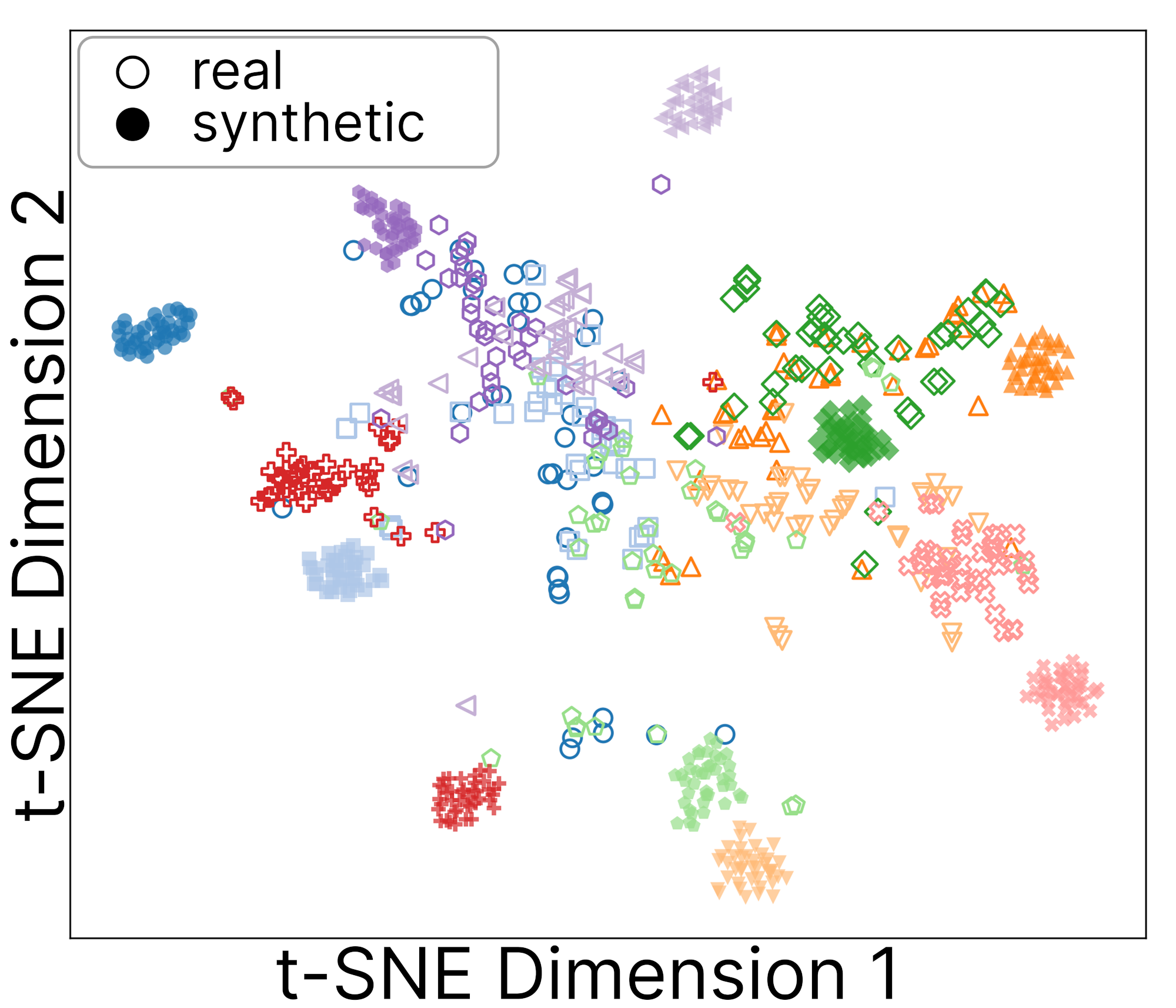}
        \caption{Fold 0}
    \end{subfigure}\hfill
    \begin{subfigure}[t]{0.32\linewidth}
        \centering
        \includegraphics[width=\linewidth]{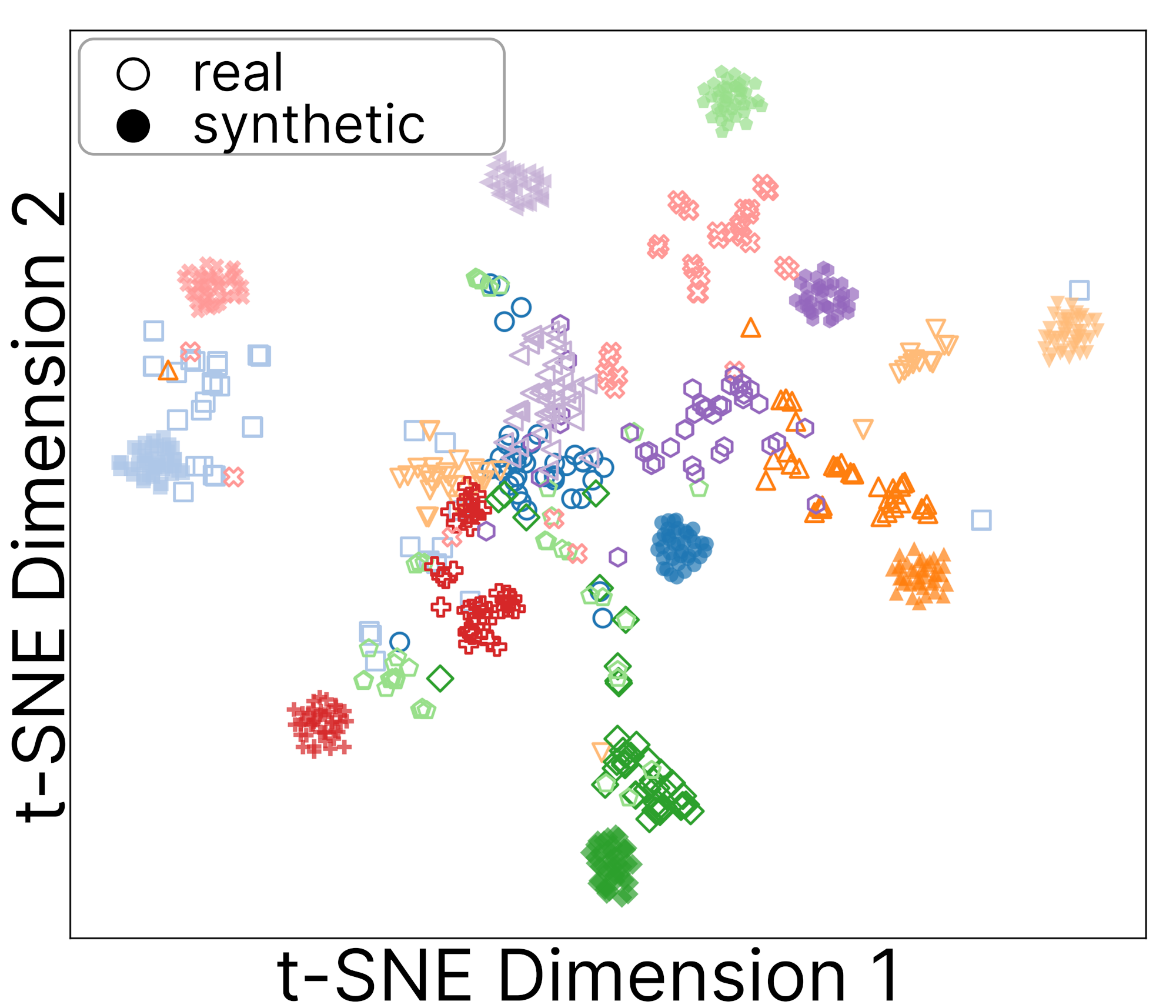}
        \caption{Fold 1}
    \end{subfigure}\hfill
    \begin{subfigure}[t]{0.32\linewidth}
        \centering
        \includegraphics[width=\linewidth]{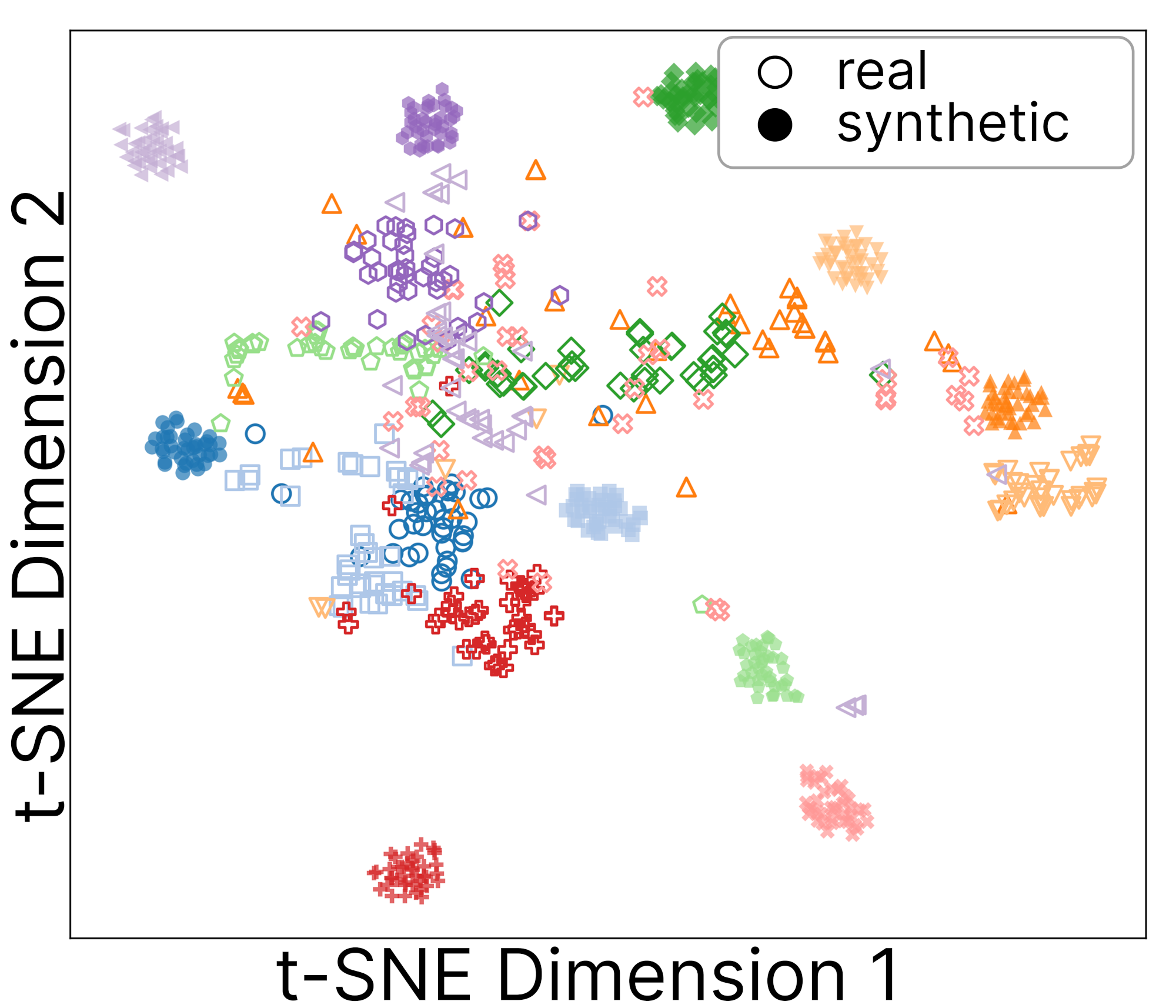}
        \caption{Fold 2}
    \end{subfigure}

    \vspace{6pt}

    \begin{subfigure}[t]{0.32\linewidth}
        \centering
        \includegraphics[width=\linewidth]{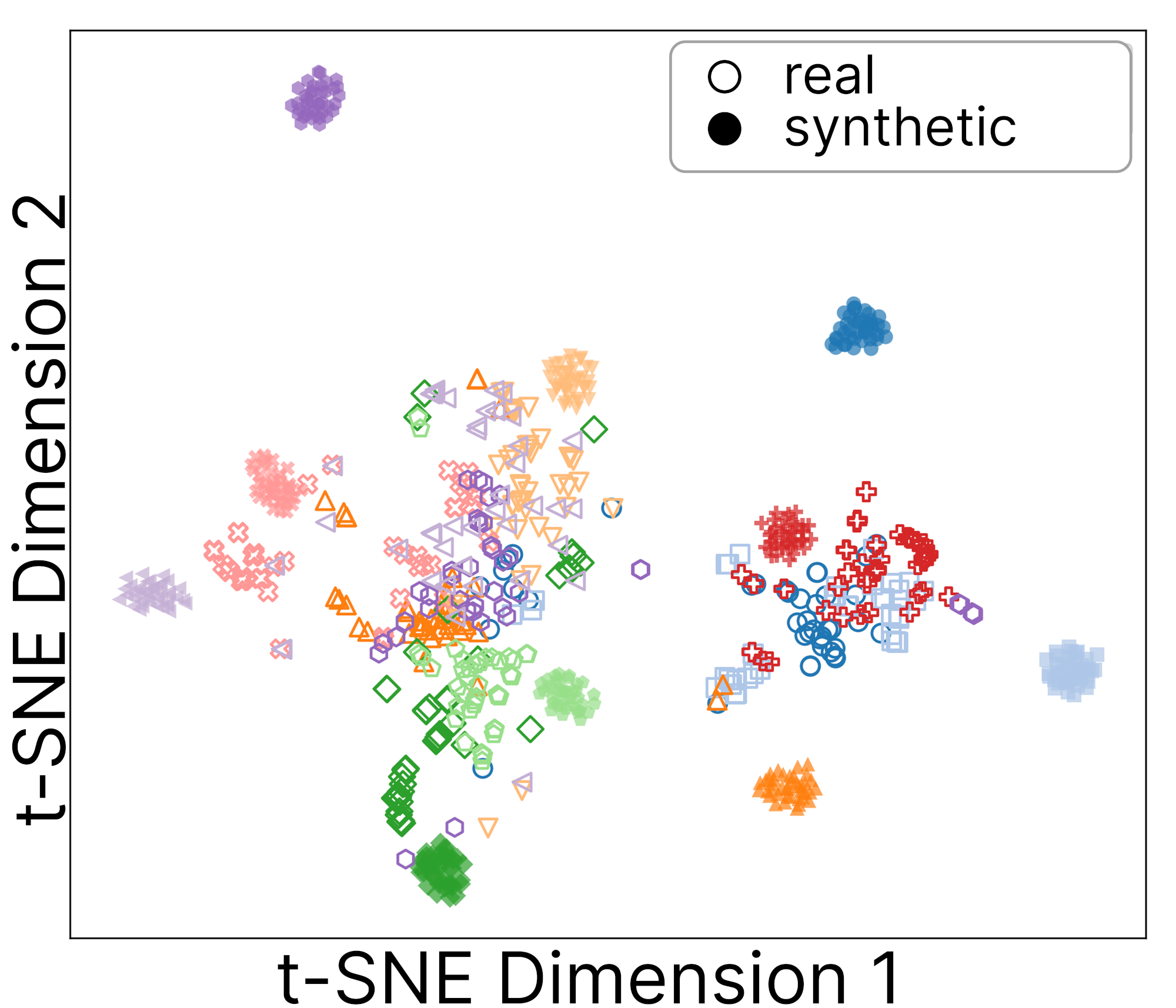}
        \caption{Fold 3}
    \end{subfigure}\hfill
    \begin{subfigure}[t]{0.32\linewidth}
        \centering
        \includegraphics[width=\linewidth]{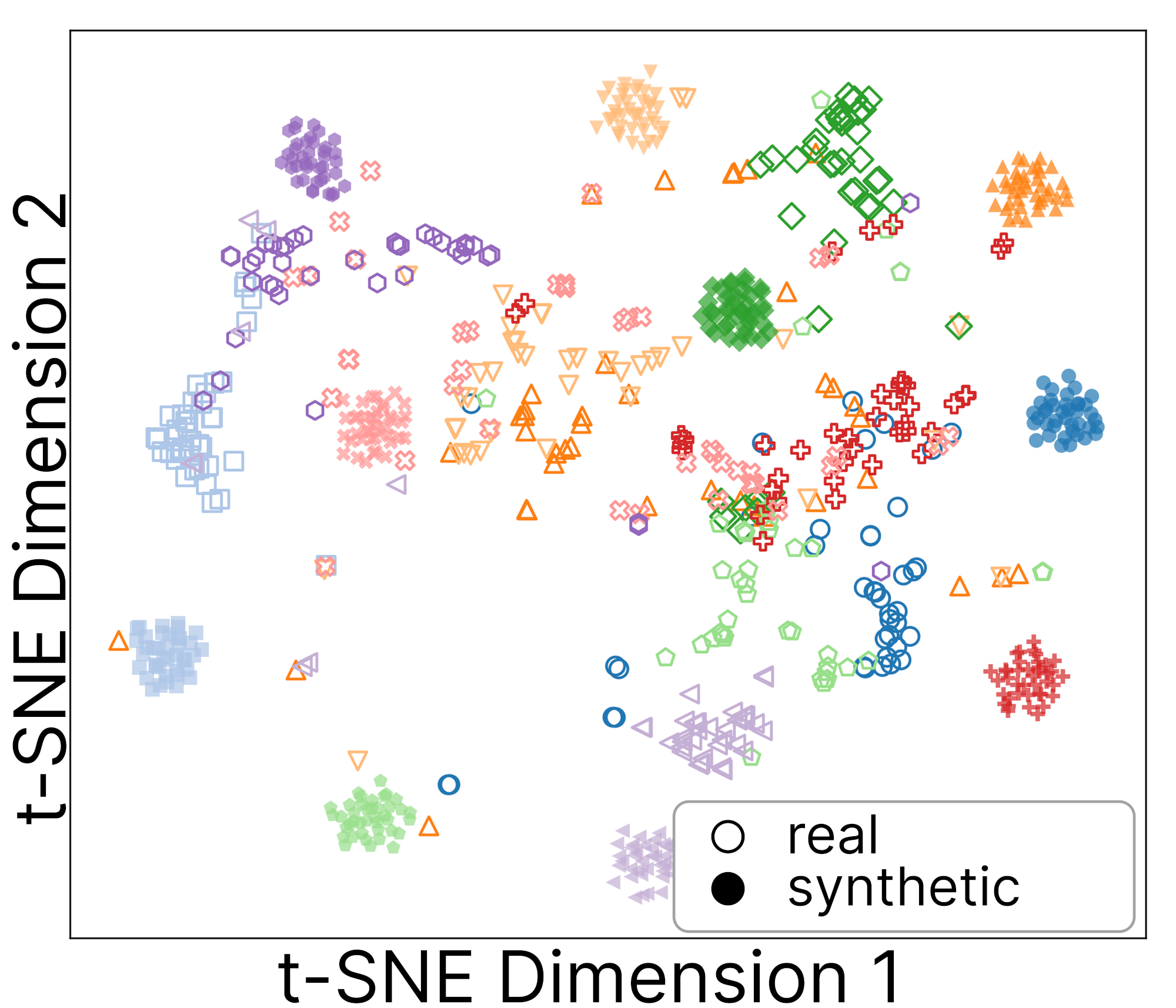}
        \caption{Test}
    \end{subfigure}\hfill
    \begin{subfigure}[t]{0.32\linewidth}
        \centering
        \mbox{}
    \end{subfigure}

    \caption{t-SNE projections of real (hollow) and CGDN-generated (filled) embeddings for the unseen classes of each ESC-50 partition. Colours and marker shapes denote classes. Synthetic embeddings form compact class-specific clusters that generally lie close to the corresponding real embeddings, although the degree of alignment varies between classes. Because t-SNE preserves local rather than global structure, distances between clusters should be interpreted with caution.}
    \label{fig:tsne}
\end{figure*}

To provide a qualitative illustration of the embeddings generated by CGDN, Figure~\ref{fig:tsne} shows t-SNE projections of synthetic unseen-class embeddings alongside the corresponding real embeddings for each ESC-50 partition.

Across all partitions, the synthetic embeddings form compact class-specific clusters rather than collapsing to a single region of the embedding space. The synthetic clusters generally appear close to the corresponding real clusters, although the degree of alignment varies considerably between classes. In some cases the two distributions are adjacent, whereas in others they are on separate sides of the t-SNE diagram.

This variability is consistent with the differences in classification accuracy observed across dataset partitions (Table~\ref{tab:results}). Better agreement between the synthetic and real distributions is expected to produce more discriminative synthetic training examples. Overall, the visualisations suggest that CGDN learns meaningful class-specific structure, while also illustrating the limitations of generating feature embeddings from semantic information alone.

As t-SNE preserves local neighbourhood structure rather than global geometry, distances between clusters should be interpreted cautiously.

\section{Discussion}

This work demonstrates that generative methods are a viable approach for zero-shot environmental sound classification. While generative zero-shot learning has been studied extensively in computer vision, comparatively little work has considered environmental audio. Across six datasets, CADA-VAE and the proposed CGDN achieve performance comparable to the strong ALE baseline, showing that feature synthesis can transfer beyond computer vision. However, the improvements reported for generative methods in image classification are not consistently observed in environmental audio, suggesting that conclusions from computer vision do not necessarily transfer across modalities.

The benchmark also provides one of the first evaluations of diffusion-inspired embedding generation for zero-shot environmental audio. Unlike previous diffusion-based zero-shot methods, which rely on large pre-trained image generation models, the DDPM and CGDN approaches operate directly in embedding space and require only task-specific embeddings. This avoids potential violations of the zero-shot assumption caused by overlap between foundation-model pre-training data and evaluation classes. Operating in embedding space also makes the methods independent of the underlying data modality and computationally inexpensive compared with raw-data generative models.

A key finding is the relationship between optimisation stability and performance. DDPM, the more complex iterative model, exhibits the highest variance across repeated runs and the weakest overall performance. In contrast, CGDN replaces iterative sampling with a single-step conditional denoising objective, resulting in substantially lower variance while achieving the highest average accuracy across datasets. Although CGDN is not significantly better than the compatibility-based ALE baseline, it significantly outperforms both LisGAN and DDPM. These results suggest that optimisation stability is an important factor for generative zero-shot learning in environmental audio, and that increasing model complexity does not necessarily improve performance.

The benchmark also highlights the importance of evaluating methods across multiple datasets and class partitions. No single approach consistently performs best across every evaluation. ALE remains competitive on several datasets, CADA-VAE performs strongly on FSC22, and CGDN achieves the highest average accuracy overall despite weaker results on UrbanSound8K and GTZAN. Performance also varies substantially across different seen and unseen class partitions within the same dataset, indicating that conclusions drawn from a single benchmark or partition may not generalise. The benchmark also establishes a reference point for future research on generative zero-shot environmental sound classification, providing a common basis for evaluating new methods across multiple audio datasets. Broader benchmarks and a better theoretical understanding of transfer between seen and unseen classes would improve the evaluation and development of future zero-shot methods.

A few limitations should be noted. The evaluation considers only the inductive zero-shot setting and does not address the generalised zero-shot setting, where seen and unseen classes are evaluated jointly. Class representations are limited to Word2Vec embeddings, which constrain the information available for transfer between semantic and feature spaces. In addition, the statistical analysis treats dataset partitions as independent observations, following common practice in classifier comparisons, although partitions within a dataset are not fully independent.

More broadly, the results highlight the limited theoretical understanding of zero-shot learning. Existing methods are largely evaluated empirically, with few guarantees regarding when knowledge can be transferred successfully between seen and unseen classes. Developing theoretical frameworks for transferability, class coverage, and expected performance remains an important direction for future research.

\section{Conclusion}

This work presented the first benchmark of generative methods for zero-shot environmental sound classification across six audio datasets. The benchmark compared compatibility-based, adversarial, variational, diffusion-based, and denoising-based approaches, including two embedding-generation methods introduced in this work.

The results show that generative methods are a viable approach for zero-shot environmental sound classification, although their advantages over established compatibility-based methods are smaller than those reported in computer vision. Across the benchmark, CGDN achieved the highest average accuracy and significantly outperformed the DDPM- and GAN-based methods, while remaining competitive with the strong ALE baseline. More broadly, the results indicate that optimisation stability is an important factor in generative zero-shot learning and that increasing model complexity does not necessarily improve performance.

Beyond evaluating existing methods, the benchmark establishes a reference point for future research in generative zero-shot environmental sound classification. Future work should expand evaluation to additional datasets and domains, investigate the generalised zero-shot setting, and develop a stronger theoretical understanding of when knowledge can be transferred successfully between seen and unseen classes.

\section*{Declarations}

\subsection*{Availability of Data and Material}

All code is available at \url{https://github.com/ysims/ZeroDiffusion}. All datasets used in this work are not owned by the authorship team, and are publicly available online.

\subsection*{Competing Interests}

The authors declare no competing interests.

\subsection*{Funding}

This research is supported by an Australian Government Research Training Program Scholarship to the first author.

\subsection*{Authors' Contributions}

All authors contributed to conceptualisation and manuscript review. Sims led methodology design, software development, analysis, and manuscript writing. Mendes and Chalup contributed AI expertise.

\subsection*{Acknowledgements}

This research is supported by an Australian Government Research Training Program Scholarship to the first author.

\bibliography{bib}
\end{document}